\documentclass[12pt]{iopart}
\usepackage[dvips]{graphicx}

\font\bsy=cmbsy12
\def\bfnabla{\vec{\hbox{\bsy\char114}}}

\font\bsy=cmbsy12
\def\enabla{\hbox{\bsy\char114}}

\begin{document}

\title[Wigner-Kirkwood expansion for semi-infinite quantum
fluids]{Wigner-Kirkwood expansion for semi-infinite quantum fluids} 

\author{L. {\v S}amaj\dag\ddag\ and B. Jancovici\dag}
\address{\dag\ Laboratoire de Physique Th\'eorique, Universit\'e
Paris-Sud, 91405 Orsay, France
(Unit\'e Mixte de Recherche no. 8627 - CNRS)}
\address{\ddag\ Institute of Physics, Slovak Academy of Sciences,
D\'ubravsk\'a cesta 9, 845 11 Bratislava, Slovakia}

\eads{\mailto{Ladislav.Samaj@savba.sk}, 
\mailto{Bernard.Jancovici@th.u-psud.fr}}

\begin{abstract}
For infinite (bulk) quantum fluids of particles interacting via pairwise
sufficiently smooth interactions, the Wigner-Kirkwood formalism provides a
semiclassical expansion of the Boltzmann density in configuration space in
even powers of the thermal de Broglie wavelength $\lambda$. 
This result permits one to generate an analogous $\lambda$-expansion
for the bulk free energy and many-body densities. 
The present paper brings a technically nontrivial generalization of 
the Wigner-Kirkwood technique to semi-infinite quantum fluids, 
constrained by a plane hard wall impenetrable to particles.
In contrast to the bulk case, the resulting Boltzmann density involves also 
position-dependent terms of type $\exp(-2x^2/\lambda^2)$ ($x$ denotes 
the distance from the wall boundary) which are non-analytic functions
of the de Broglie wavelength $\lambda$. 
Under some condition, the analyticity in $\lambda$ is restored by 
integrating the Boltzmann density over configuration space; 
however, in contrast to the bulk free energy, the semiclassical expansion of 
the surface part of the free energy (surface tension) contains odd powers 
of $\lambda$, too.
Explicit expressions for the leading quantum corrections in the presence
of the boundary are given for the one-body and two-body densities.
As model systems for explicit calculations, we use Coulomb fluids, in
particular the one-component plasma defined in the $\nu$-dimensional (integer
$\nu\ge 2$) space.   
\end{abstract}


\medskip


\maketitle

\eqnobysec

\section{Introduction}
In this paper, we consider a quantum system of $N$ identical particles 
$j=1,2,\ldots,N$ of mass $m$ which position vectors 
${\bf r}_1,{\bf r}_2,\ldots,{\bf r}_N$ are confined to a region $\Lambda$ 
of the $\nu$-dimensional space $R^{\nu}$.
For the sake of brevity, we denote the $\nu N$-dimensional position
vector in configuration space by 
$\vec{{\bf r}} = ({\bf r}_1,{\bf r}_2,\ldots,{\bf r}_N)$ and the corresponding 
gradient by $\bfnabla = (\enabla_1,\enabla_2,\ldots,\enabla_N)$.
The Hamiltonian of the particles (in the absence of a magnetic field) is
\begin{equation} \label{1.1}
H = \frac{1}{2 m} \left( - {\rm i}\hbar\bfnabla \right)^2
+ V(\vec{{\bf r}}) ,
\end{equation}
where $\hbar$ is Planck's constant and the scalar potential $V(\vec{{\bf r}})$ 
includes pairwise particle interactions 
$\sum_{j<k} v(\vert{\bf r}_j-{\bf r}_k\vert)$
plus perhaps one-particle interactions $\sum_j u({\bf r}_j)$ 
with some external potential $u({\bf r})$.
The equilibrium statistical mechanics of the particle system 
is studied in the canonical ensemble at the temperature $T$ 
(or, alternatively, the inverse temperature $\beta=1/k_{\rm B}T$ 
with $k_{\rm B}$ being Boltzmann's constant).
In the classical regime, the Hamiltonian (\ref{1.1}) is replaced by 
$H=\vec{\bf p}^2/(2m) + V(\vec{\bf r})$, where 
$\vec{\bf p}=({\bf p}_1,{\bf p}_2,\ldots,{\bf p}_N)$ is the vector
in the classical momentum space.
The integration of the Boltzmann density $\exp(-\beta H)$ over
the momentum variables results in a trivial constant, so that the
classical Boltzmann density in configuration $\vec{\bf r}$-space 
$\propto \exp[-\beta V(\vec{\bf r})]$. 
Quantum effects are related to the thermal de Broglie wavelength
$\lambda = \hbar (\beta/m)^{1/2}$. 
A {\em nearly classical} regime of present interest is defined 
as being such that the dimensionless parameter $\lambda/l$ ($l$ being 
a typical microscopic length of the classical particle system, 
proportional to the mean interparticle distance) is sufficiently small.

For a nearly classical bulk system in an infinite $\nu$-dimensional 
space $\Lambda = R^{\nu}$, the Boltzmann density in configuration space 
$\vec{\bf r}$ can be expanded in powers of $\lambda^2$ (odd powers
of $\lambda$ vanish) within the well-known 
Wigner-Kirkwood expansion \cite{Wigner,Kirkwood}. 
The coefficients of the $\lambda^2$-expansion of the partition function,
which is given by the integration of the Boltzmann density over 
configuration space $\vec{\bf r}$, are expressible in terms of
the classical averages of gradient operations applied to $V(\vec{\bf r})$
and of their products.

The applicability of the Wigner-Kirkwood expansion scheme depends
on the form of the interaction potential $v(r)$.
For certain potentials, the expansion in $\lambda^2$ is analytic
in the sense of holomorphy.
There are potentials for which the Wigner-Kirkwood expansion
is a limited Taylor expansion in $\lambda^2$: divergence of 
an expansion coefficient indicates a loss of analyticity of 
the corresponding and higher-order quantum corrections.
Like for instance, for the 3D exponential, screened Coulomb and 
square barrier potentials, the leading expansion term was found to have 
the non-analytic form $\sqrt{\lambda^2}$ \cite{DeWitt62a}. 
In the case of an inverse-power-law repulsive potential $v(r)=v_0 (a/r)^n$ 
with the exponent $n$ from the range $1<n<\infty$, the Wigner-Kirkwood 
expansion turns out to be analytic in $\lambda^2$ in the holomorphic sense 
\cite{DeWitt62a}.
In the hard-core limit $n\to\infty$, the Wigner-Kirkwood expansion fails
and one has once more the nonanalyticity of type $\sqrt{\lambda^2}$,
as was shown in numerous analytic studies (see, e.g., 
\cite{Hill,Jancovici69,Pisani,Mason}).
  
Much effort has been devoted to the calculation of quantum corrections
for 3D Coulomb fluids of $\pm e$ charges, in the high-temperature regime 
described adequately by the classical Debye-H\"uckel theory.
Quantum corrections of the bulk equation of state were derived and
analyzed in Ref. \cite{DeWitt62b}.
As concerns the two-body density, its leading quantum correction
exhibits at large distances an inverse-power-law decay in contrast with the
classical Debye-H\"uckel exponential screening (see the review \cite{Brydges}).

To our knowledge, there does not exist in the literature a boundary
version of the Wigner-Kirkwood expansion applicable to fluids in domains 
$\Lambda$ surrounded by hard walls impenetrable to particles
(or, equivalently, there is an infinite external potential outside
the domain $\Lambda$).
Surface properties of quantum particle systems have been investigated only
in the presence of some smooth confining external-potential barriers
(see, e.g., \cite{Wang,Centelles}).
The sole exceptions are represented by semiclassical studies of 
3D quantum Coulomb fluids in the high-temperature Debye-H\"uckel limit
for which one has the explicit results for the density profile near a plain 
hard wall \cite{Aqua04} and for the large-distance tail of charge
correlations along a plain hard wall or a conducting wall \cite{Aqua00}.
In these explicit results, quantum corrections linear in $\lambda$
together with non-analytic terms of type $\exp(-2x^2/\lambda^2)$
($x$ denotes the distance from the boundary) appear; the latter
are linked to the propagator of a free particle \cite{Kleinert}.

The present paper brings a technically nontrivial generalization of 
the Wigner-Kirkwood technique to semi-infinite quantum fluids, 
constrained by a plane hard wall.
In contrast to the bulk case, the resulting Boltzmann density also involves
position-dependent terms which are non-analytic functions of the de Broglie
thermal wavelength $\lambda$. 
Under some condition about the classical density profile,
the analyticity in $\lambda$ is restored by integrating the Boltzmann
density over configuration space; however, in contrast to the bulk
free energy, the semiclassical expansion of the surface part of 
the free energy (surface tension) contains odd powers of $\lambda$, too.
Explicit expressions for the leading quantum corrections in the presence
of the boundary are given for one-body and two-body densities.

As a model system for explicit calculations, we shall use a special kind of 
Coulomb fluid, namely the one-component plasma (jellium) made of mobile
pointlike charges $e$ neutralized by a uniform oppositely charged fixed 
background, in the $\nu$-dimensional space (because of some physical 
nontrivialities in 1D, we shall restrict ourselves to dimensions $\nu\ge 2$).
In the infinite space $\Lambda=R^{\nu}$, or when the hard walls surrounding
the confining domain $\Lambda$ do not induce image charges (plain hard walls),
the translationally invariant Coulomb interaction potential of two $e$-charges 
$v(r)$ with $r=\vert {\bf r}_j-{\bf r}_k \vert$ is defined as the solution 
of the Poisson equation $\enabla^2 v = - s_{\nu} e^2 \delta({\bf r})$, where
$s_{\nu} = 2 \pi^{\nu/2}/\Gamma(\nu/2)$ ($\Gamma$ stands for the Gamma 
function) is the surface area of the $\nu$-dimensional unit sphere.
The resulting Coulomb interaction potential is logarithmic in $\nu=2$ 
dimensions and of type $e^2 r^{2-\nu}/(\nu-2)$ in $\nu\ge 3$ dimensions. 
The present definition of the Coulomb potential maintains many generic 
properties of ``real'' 3D Coulomb systems, like screening 
and the corresponding sum rules (see the review \cite{Martin}).
The total potential $V(\vec{\bf r})$ then satisfies for each of the particle
coordinates the differential equation 
\begin{equation} \label{1.2}
\enabla_j^2 V(\vec{\bf r}) = - s_{\nu} e^2 \sum_{k=1\atop(k\ne j)}^N 
\delta({\bf r}_j-{\bf r}_k) + s_{\nu} e^2 n ,
\quad j=1,2,\ldots,N .
\end{equation}
Here, the second term on the rhs comes from the particle-background 
interaction and $n=N/\vert\Lambda\vert$ is the mean number density of 
the mobile charges; the background-background interaction constant 
contributes to the partition function. 

The paper is organized as follows.
In section 2, the method of the bulk Wigner-Kirkwood expansion is briefly 
reviewed in a format which can be relatively simply extended to the
boundary case.
The following sections are devoted to semi-infinite quantum systems
constrained by a plane hard wall. 
In section 3, we propose a method for constructing 
the expansion of the Boltzmann density in configuration 
$\vec{\bf r}$-space.
This expansion is subsequently used in section 4 for the generation
of the expansion for statistical quantities, such as
the partition function and the corresponding surface part of the free energy, 
one-body and two-body densities.
Explicit calculations for Coulomb models with known classical statistical
quantities are presented in section 5.
A few concluding remarks are given in section 6.

\section{Wigner-Kirkwood expansion in infinite space}
Let the particle system be in the infinite space $\Lambda=R^{\nu}$.
In this so-called ``bulk'' regime, its equilibrium quantities in the nearly 
classical regime can be expanded in powers of $\hbar^2$ within the standard 
Wigner-Kirkwood expansion \cite{Wigner,Kirkwood} which {\em neglects} fermion 
or boson exchange effects between quantum particles.
In this section, we review shortly the derivation of this expansion by the 
Laplace transform method \cite{Hill,Alastuey} which is suitable for our next 
purposes.

\subsection{Boltzmann density}
The Boltzmann density $B_{\beta}$ in configuration space 
$\vec{\bf r}$ can be formally written in the basis of plane waves as 
a $\nu N$-dimensional integral defined in an infinite domain $R^{\nu}$:
\begin{equation} \label{2.1}
B_{\beta}(\vec{\bf r}) \equiv 
\langle \vec{{\bf r}} \vert {\rm e}^{-\beta H} \vert \vec{{\bf r}} \rangle
= \int \frac{{\rm d}\vec{\bf p}}{(2\pi\hbar)^{\nu N}}
{\rm e}^{-({\rm i}/\hbar) \vec{{\bf p}}\cdot\vec{{\bf r}}} {\rm e}^{-\beta H}
{\rm e}^{({\rm i}/\hbar) \vec{{\bf p}}\cdot\vec{{\bf r}}} , 
\end{equation}
where $\vec{{\bf p}}=({\bf p}_1,{\bf p}_2,\ldots,{\bf p}_N)$ is the
$\nu N$-dimensional momentum vector.
Instead of considering ${\rm e}^{-\beta H}$ in the integration (\ref{2.1}),
we take the Laplace transform of this operator with respect to 
the inverse temperature $\beta$,
\begin{equation} \label{2.2}
\frac{1}{H + z} = \int_0^{\infty} {\rm d}\beta\, {\rm e}^{-\beta z}
{\rm e}^{-\beta H} ,
\end{equation}
and concentrate on the generation of an expansion in powers of $\hbar$ for
\begin{equation} \label{2.3}
{\rm e}^{-({\rm i}/\hbar) \vec{{\bf p}}\cdot\vec{{\bf r}}} \frac{1}{H+z}
{\rm e}^{({\rm i}/\hbar) \vec{{\bf p}}\cdot\vec{{\bf r}}} . 
\end{equation}
Let us first rewrite $H+z$ as
\begin{equation} \label{2.4}
H + z = D + {\cal O} ,
\end{equation}
where $D$ is a $c$-number
\begin{equation} \label{2.5}
D = \frac{1}{2 m} \vec{\bf p}^2  + V(\vec{{\bf r}}) + z 
\end{equation}
and ${\cal O}$ an operator
\begin{equation} \label{2.6}
{\cal O} = \frac{1}{2 m} \left( -{\rm i}\hbar\bfnabla \right)^2
- \frac{1}{2 m} \vec{\bf p}^2 ,
\end{equation}
and then expand
\begin{equation} \label{2.7}
\frac{1}{H + z} = \frac{1}{D} - \frac{1}{D} {\cal O} \frac{1}{D} 
+ \frac{1}{D} {\cal O} \frac{1}{D} {\cal O} \frac{1}{D} - \cdots .
\end{equation}
It can be verified that for any function $f(\vec{{\bf r}})$ 
the operator ${\cal O}$ acts as follows
\begin{equation} \label{2.8}
{\cal O} \left[ f(\vec{{\bf r}}) {\rm e}^{({\rm i}/\hbar) 
\vec{{\bf p}}\cdot\vec{{\bf r}}} \right] 
= - {\rm e}^{({\rm i}/\hbar) \vec{{\bf p}}\cdot\vec{{\bf r}}} \left[
\frac{{\rm i}\hbar}{m} \vec{\bf p}\cdot\bfnabla + \frac{\hbar^2}{2 m}
\bfnabla^2 \right] f(\vec{{\bf r}}) .
\end{equation}
One thus finds that
\begin{equation} \label{2.9}
{\rm e}^{-({\rm i}/\hbar) \vec{{\bf p}}\cdot \vec{{\bf r}}}\, \frac{1}{H+z}\,
{\rm e}^{({\rm i}/\hbar) \vec{{\bf p}}\cdot \vec{{\bf r}}}
= \frac{1}{D} \sum_{n=0}^{\infty} \left\{
\left[ \frac{{\rm i}\hbar}{m} \vec{\bf p}\cdot\bfnabla
+ \frac{\hbar^2}{2 m} \bfnabla^2 \right] \frac{1}{D} \right\}^n .
\end{equation}

The term of order $n$ in (\ref{2.9})
\begin{equation} \label{2.10}
\frac{1}{D} \left\{ \left[ \frac{{\rm i}\hbar}{m} \vec{\bf p}\cdot\bfnabla
+ \frac{\hbar^2}{2 m} \bfnabla^2 \right] \frac{1}{D} \right\}^n 
\end{equation}
is a polynomial in $\hbar$ containing powers in the range from $\hbar^n$ 
up to $\hbar^{2n}$.
This means that the truncation of the series (\ref{2.9}) at some given order 
$n$ provides all $\hbar$-terms up to order $\hbar^n$ 
(or, equivalently, all $\lambda$-terms up to order $\lambda^n$).
By carrying out all differentiations in (\ref{2.10}), 
taking subsequently the inverse Laplace-transforms using the formula
\begin{eqnarray}
\frac{1}{D^j} & = & \frac{(-1)^{j-1}}{(j-1)!} 
\frac{\partial^{j-1}}{\partial z^{j-1}} 
\frac{1}{z + \vec{\bf p}^2/(2m) + V(\vec{{\bf r}})} \nonumber \\
& = & \int_0^{\infty} {\rm d}\beta\, {\rm e}^{-\beta z}
\frac{1}{(j-1)!} \beta^{j-1} 
{\rm e}^{-\beta[\vec{\bf p}^2/(2m) + V(\vec{{\bf r}})]} \label{2.11}
\end{eqnarray}
$(j=1,2,\ldots)$, and finally integrating on the momentum variables 
$\vec{\bf p}$, the Boltzmann density in configuration space (\ref{2.1}) 
is obtained as the series
\begin{equation} \label{2.12}
\langle \vec{{\bf r}} \vert {\rm e}^{-\beta H} \vert \vec{{\bf r}} \rangle
= \sum_{n=0}^{\infty} B_{\beta}^{(n)}(\vec{\bf r}) ,
\end{equation}
where $B_{\beta}^{(n)}(\vec{\bf r})$ denotes the contribution
implied by the $n$th term (\ref{2.10}).
In particular,
\numparts
\begin{eqnarray}
\fl B_{\beta}^{(0)}(\vec{\bf r}) & = &
\frac{1}{(\sqrt{2\pi} \lambda)^{\nu N}} {\rm e}^{-\beta V} , \label{2.13a} \\
\fl B_{\beta}^{(1)}(\vec{\bf r}) & = &
\frac{1}{(\sqrt{2\pi} \lambda)^{\nu N}} {\rm e}^{-\beta V}
\, \lambda^2 \left[ - \frac{\beta}{4} \bfnabla^2 V
+ \frac{\beta^2}{6} \left(\bfnabla V\right)^2 \right] , \label{2.13b} \\
\fl B_{\beta}^{(2)}(\vec{\bf r}) & = &
\frac{1}{(\sqrt{2\pi} \lambda)^{\nu N}} {\rm e}^{-\beta V}
\left\{  \lambda^2 \left[ \frac{\beta}{6} \bfnabla^2 V
- \frac{\beta^2}{8} \left( \bfnabla V\right)^2 \right] 
+ \lambda^4 \left[ 
-\frac{\beta}{24} \left(\bfnabla^2 \right)^2 V
\right. \right. \nonumber \\ 
\fl & & \left. \left. 
+\frac{\beta^2}{16} \bfnabla V \cdot \bfnabla\left( \bfnabla^2 V\right)
+\frac{\beta^2}{48} \bfnabla^2\left( \bfnabla V\right)^2 
+\frac{\beta^2}{32} \left(\bfnabla^2 V\right)^2 
\right. \right. \label{2.13c} \\ 
\fl & & \left. \left.
-\frac{\beta^3}{30} \bfnabla V \cdot \bfnabla\left( \bfnabla V\right)^2
-\frac{\beta^3}{24} \left(\bfnabla V\right)^2 \bfnabla^2 V
+ \frac{\beta^4}{72}\left(\bfnabla V\right)^4 \right] \right\} , \nonumber
\end{eqnarray}
\endnumparts
etc.
We conclude that the quantum Boltzmann density in configuration space 
is given, to order $\lambda^2$, by
\begin{eqnarray} 
\fl \langle \vec{{\bf r}} \vert {\rm e}^{-\beta H} \vert \vec{{\bf r}} \rangle
& = & \frac{1}{(\sqrt{2\pi} \lambda)^{\nu N}} {\rm e}^{-\beta V}
\left\{ 1 + \lambda^2 \left[ - \frac{\beta}{12} \bfnabla^2 V
+ \frac{\beta^2}{24} \left( \bfnabla V \right)^2 \right] 
+ O(\lambda^4) \right\}  \nonumber \\
\fl & = & \frac{1}{(\sqrt{2\pi} \lambda)^{\nu N}} \left\{
{\rm e}^{-\beta V} \left[ 1 - \frac{\lambda^2\beta}{24} \bfnabla^2 V \right]
+ \frac{\lambda^2}{24} \bfnabla^2 {\rm e}^{-\beta V}
+ O(\lambda^4) \right\} . \label{2.14}
\end{eqnarray}
When integrated over the coordinates of particle $j$, the corresponding
term in $\sum_{j=1}^N \enabla_j^2 \exp(-\beta V)$ gives no volume
contribution.
Notice that odd powers of $\lambda$ do not appear in the Wigner-Kirkwood 
expansion, so that the bulk Boltzmann factor, when considered without
the trivial prefactor $(\sqrt{2\pi}\lambda)^{-\nu N}$, is invariant 
with respect to the transformation ${\rm i}\hbar \to -{\rm i}\hbar$.  
This is due to the fact that the integration of terms with odd powers of
momentum components over the whole corresponding axis is 0. 

\subsection{Statistical quantities}
According to the standard formalism of statistical quantum mechanics,
the partition function of the $N$-particle fluid
(with ignored exchange effects) is given by the integration
of the Boltzmann density over configuration space:
\begin{equation} \label{2.15}
Z_{\rm qu} = \frac{1}{N!} \int_{\Lambda} {\rm d}\vec{\bf r}\,
\left\langle \vec{\bf r} \vert {\rm e}^{-\beta H} \vert \vec{\bf r}
\right\rangle .
\end{equation}
For expressing macroscopic physical quantities, one defines the
quantum average of a function $f(\vec{\bf r})$ as follows
\begin{equation} \label{2.16}
\left\langle f \right\rangle_{\rm qu} = 
\frac{1}{Z_{\rm qu}} \frac{1}{N!} 
\int_{\Lambda} {\rm d}\vec{\bf r}\, 
\left\langle \vec{\bf r} \vert {\rm e}^{-\beta H} \vert \vec{\bf r}
\right\rangle f(\vec{\bf r}) .
\end{equation}
At the one-particle level, one introduces the particle density
\begin{equation} \label{2.17}
n_{\rm qu}({\bf r}) = \Bigg\langle
\sum_{j=1}^N \delta({\bf r}-{\bf r}_j) \Bigg\rangle_{\rm qu} ,
\end{equation}
At the two-particle level, the two-body density is given by
\begin{equation} \label{2.18}
n_{\rm qu}^{(2)}({\bf r},{\bf r}') = \Bigg\langle
\sum_{j,k=1\atop (j\ne k)}^N 
\delta({\bf r}-{\bf r}_j) \delta({\bf r}'-{\bf r}_k) \Bigg\rangle_{\rm qu} .
\end{equation}
It will be useful to consider also the truncated two-body density
\begin{equation} \label{2.19}
n_{\rm qu}^{(2){\rm T}}({\bf r},{\bf r}') = n_{\rm qu}^{(2)}({\bf r},{\bf r}')
- n_{\rm qu}({\bf r}) n_{\rm qu}({\bf r}') 
\end{equation}
vanishing at asymptotically large distances
$\vert {\bf r}-{\bf r}'\vert\to\infty$,
the pair distribution function $g_{\rm qu}({\bf r},{\bf r}') =  
n_{\rm qu}^{(2)}({\bf r},{\bf r}')/[n_{\rm qu}({\bf r}) n_{\rm qu}({\bf r}')]$
and the pair correlation function  
$h_{\rm qu}({\bf r},{\bf r}') =  g_{\rm qu}({\bf r},{\bf r}') - 1$.
The classical partition function and the classical average of a function
$f(\vec{\bf r})$ are defined as follows
\begin{eqnarray}
Z & = & \frac{1}{N!} \int_{\Lambda} 
\frac{{\rm d}\vec{\bf r}}{(\sqrt{2\pi}\lambda)^{\nu N}}\,
{\rm e}^{-\beta V(\vec{\bf r})} , \label{2.20} \\
\left\langle f \right\rangle & = &  \frac{1}{Z} \frac{1}{N!} 
\int_{\Lambda} \frac{{\rm d}\vec{\bf r}}{(\sqrt{2\pi}\lambda)^{\nu N}}\,
{\rm e}^{-\beta V(\vec{\bf r})} f(\vec{\bf r}) . \label{2.21}
\end{eqnarray}
According to our convention, the classical values of statistical quantities
will be written without a subscript, like 
$n({\bf r})$, $n^{(2)}({\bf r},{\bf r}')$, etc.

Substituting the $\lambda$-expansion of the Boltzmann density (\ref{2.14})
into formula (\ref{2.15}), the quantum partition function takes the
expansion form
\begin{equation} \label{2.22}
Z_{\rm qu} = Z \left[ 1 - 
\lambda^2 \frac{\beta}{24}\langle \bfnabla^2 V\rangle 
+ O(\lambda^4) \right] .
\end{equation}
It is important to note that the general applicability of the 
Wigner-Kirkwood expansion scheme depends on the particular form of
the interaction potential $V$; a divergence of the classical average 
$\langle \bfnabla^2 V\rangle$ indicates a non-analytic $\lambda^2$-expansion 
of the free energy.

For the model system of the one-component plasma in $\nu$ dimensions,
the summation over the particle index $j$ of the set of Poisson equations
(\ref{1.2}) leads to
\begin{equation} \label{2.23}
\bfnabla^2 V(\vec{\bf r}) = - s_{\nu} e^2 \sum_{j,k=1\atop (j\ne k)}^N
\delta({\bf r}_j-{\bf r}_k) + s_{\nu} N e^2 n .
\end{equation}
The first term on the rhs of (\ref{2.23}) does not give any contribution
to $\langle \bfnabla^2 V\rangle$, because it is weighted by
a classical Boltzmann factor which vanishes at zero interparticle distance.
The replacement of $\bfnabla^2 V$ in equation (\ref{2.22}) by 
the particles-background term $s_{\nu} N e^2 n$ leads to  
\begin{equation} \label{2.24}
Z_{\rm qu} = Z \left[ 1 - 
\lambda^2 N \frac{\beta s_{\nu} e^2 n}{24} + O(\lambda^4) \right] .
\end{equation}
For the free energy $F_{\rm qu}$ defined by the relation
$-\beta F_{\rm qu} = \ln Z_{\rm qu}$, one gets from (\ref{2.24})
the leading $\lambda^2$-correction term
\begin{equation} \label{2.25}
F_{\rm qu} = F + \lambda^2 N \frac{s_{\nu} e^2 n}{24} + O(\lambda^4) ,
\end{equation}
which is, as one expects in the bulk regime, extensive in the number 
of particles $N$.
Corrections to the free energy of higher order in $\lambda$ are standardly
obtained from the partition function with higher-order $\lambda$-terms
by using the cumulant expansion (see, e.g., Ref. \cite{Alastuey}).

The Wigner-Kirkwood formalism ignores the exchange effects. However, for the
one-component plasma, they are exponentially small \cite{Jancovici78}.

The $\lambda$-expansion of the particle density (\ref{2.17}) 
for the one-component plasma reads 
\begin{equation} \label{2.26}
n_{\rm qu}({\bf r}) = \left[ 1 + \frac{\lambda^2}{24} \enabla^2
+ O(\lambda^4) \right] n({\bf r}) .
\end{equation}  
In the bulk regime with the homogeneous classical density of particles 
$n({\bf r}) = n$, one obtains the trivial result $n_{\rm qu}({\bf r}) = n$.

The radial distribution function of the bulk one-component plasma
has a $\lambda$-expansion of the form \cite{Jancovici76}
\begin{equation} \label{2.27}
g_{\rm qu}(r) = \left[ 1 + \frac{\lambda^2}{12} \enabla^2
+ O(\lambda^4) \right] g(r) .
\end{equation}
In the short-distance $r\to 0$ limit, $g(r)$ is proportional to the
Boltzmann factor of the bare Coulomb potential, which causes 
the breaking down of the Wigner-Kirkwood expansion scheme for
too small values of $r$ \cite{Jancovici76}.

\section{Half-space geometry: Boltzmann density}
Let the particles be constrained to the half-space $\Lambda$
defined by Cartesian coordinates ${\bf r} = (x,{\bf r}^{\perp})$,
where $x>0$ and ${\bf r}^{\perp}\in R^{\nu-1}$ denotes the set of 
$(\nu-1)$ unbounded spatial coordinates normal to $x$.
The hard wall in half-space $x<0$ is considered to be impenetrable 
to particles.
We group the components of the $\nu N$-dimensional vector in configuration
space $\vec{{\bf r}}=({\bf r}_1,{\bf r}_2,\ldots,{\bf r}_N)$ into two another
vectors $\vec{x}=(x_1,x_2,\ldots,x_N)$ and $\vec{{\bf r}}^{\perp} =
({\bf r}_1^{\perp},{\bf r}_2^{\perp},\ldots,{\bf r}_N^{\perp})$;
the corresponding momentum vectors generated from
$\vec{{\bf p}}=({\bf p}_1,{\bf p}_2,\ldots,{\bf p}_N)$ are 
$\vec{p}^x = (p_1^x,p_2^x,\ldots,p_N^x)$ and $\vec{{\bf p}}^{\perp} =
({\bf p}_1^{\perp},{\bf p}_2^{\perp},\ldots,{\bf p}_N^{\perp})$.
The Hamiltonian is still of the form (\ref{1.1}), where the interaction 
potential $V(\vec{\bf r})$ may include also additional effective 
one-body and two-body terms induced by the hard wall, e.g., the Coulomb 
energy of images in the case of charged particles near a dielectric wall.

Due to the presence of the impenetrable hard wall, the wave function 
of the particle system has to vanish as soon as one of the particles 
lies at the wall.
The one-body Boltzmann density in configuration space can be thus written in 
the basis of stationary plane waves with zero Dirichlet boundary condition 
at $x=0$:
\begin{equation} \label{3.1}
B_{\beta}(\vec{\bf r}) \equiv
\langle \vec{{\bf r}} \vert {\rm e}^{-\beta H} \vert \vec{{\bf r}} \rangle
= \int_0^{\infty} \frac{{\rm d}\vec{p}^x}{(\pi\hbar)^N}
\int \frac{{\rm d}\vec{\bf p}^{\perp}}{(2\pi\hbar)^{(\nu-1)N}}
\psi_{\vec{\bf p}}^*(\vec{{\bf r}})\, {\rm e}^{-\beta H}\, 
\psi_{\vec{\bf p}}(\vec{{\bf r}}) ,
\end{equation} 
where
\begin{equation} \label{3.2}
\psi_{\vec{\bf p}}(\vec{{\bf r}}) =
{\rm e}^{({\rm i}/\hbar)\vec{{\bf p}}^{\perp}\cdot\vec{{\bf r}}^{\perp}} 
\prod_{j=1}^N \left[ \sqrt{2} \sin\left( \frac{p_j^x x_j}{\hbar} \right) 
\right] .
\end{equation}
To generate the $\hbar$-expansion of the matrix element (\ref{3.1}),
we introduces the Laplace transform of the operator ${\rm e}^{-\beta H}$ 
and consider
\begin{equation} \label{3.3}
\psi_{\vec{\bf p}}^*(\vec{{\bf r}})\, \frac{1}{H+z}\, 
\psi_{\vec{\bf p}}(\vec{{\bf r}}) .
\end{equation}
One can proceed formally as in the previous bulk case, up to the expansion 
(\ref{2.7}).
Afterwards, one needs the following generalization of the operator equation 
(\ref{2.8}):
\begin{equation} \label{3.4}
{\cal O} \left[ f(\vec{{\bf r}}) {\rm e}^{({\rm i}/\hbar) 
\vec{{\bf p}}'\cdot\vec{{\bf r}}}
\right] = - {\rm e}^{({\rm i}/\hbar) \vec{{\bf p}}'\cdot\vec{{\bf r}}} \left[
\frac{{\rm i}\hbar}{m} \vec{\bf p}'\cdot\bfnabla + \frac{\hbar^2}{2 m}
\bfnabla^2 \right] f(\vec{{\bf r}}) \,,
\end{equation}
which is valid for any $\nu N$-dimensional momentum vector $\vec{\bf p}'$
such that $\vert \vec{\bf p}'\vert = \vert \vec{\bf p}\vert$.
As a consequence, one finds from (\ref{2.7}) that
\begin{equation} \label{3.5} 
\frac{1}{H+z}\, {\rm e}^{({\rm i}/\hbar) \vec{{\bf p}}'\cdot \vec{{\bf r}}}
=  {\rm e}^{({\rm i}/\hbar) \vec{{\bf p}}'\cdot \vec{{\bf r}}}
\frac{1}{D} \sum_{n=0}^{\infty} \left\{
\left[ \frac{{\rm i}\hbar}{m} \vec{\bf p}'\cdot\bfnabla
+ \frac{\hbar^2}{2 m} \bfnabla^2 \right] \frac{1}{D} \right\}^n . 
\end{equation}
To make use of this operator formula, we consider an exponential
representation of the function $\psi_{\vec{\bf p}}(\vec{{\bf r}})$ 
given by equation (\ref{3.2}):
\begin{eqnarray} 
\psi_{\vec{\bf p}}(\vec{{\bf r}}) & = & 
{\rm e}^{({\rm i}/\hbar) \vec{{\bf p}}^{\perp}\cdot \vec{{\bf r}}^{\perp}}
\frac{1}{(\sqrt{2}{\rm i})^N} \prod_{j=1}^N 
\sum_{\sigma_j=\pm 1} \sigma_j {\rm e}^{({\rm i}/\hbar) \sigma_j p_j^x x_j} 
\nonumber \\ & = & 
\frac{1}{(\sqrt{2}{\rm i})^N} \sum_{\{ \vec{\sigma}\}} 
\left( \prod_{j=1}^N \sigma_j \right) 
{\rm e}^{({\rm i}/\hbar)\vec{\bf p}(\vec{\sigma})\cdot\vec{\bf r}} .
\label{3.6}
\end{eqnarray}
Here, each of the $\sigma$-components of the ``state'' vector 
$\vec{\sigma} = (\sigma_1,\sigma_2,\ldots,\sigma_N)$ can take one of 
the values $\pm 1$ (the sign determines the direction of the plane wave 
propagation) and the summation goes over all possible $2^N$ 
$\vec\sigma$-configurations.
For a given state vector $\vec\sigma$, we have redefined the particle momenta 
${\bf p}_j(\sigma_j)=(\sigma_j p_j^x,{\bf p}_j^{\perp})$  
$(j=1,2,\ldots,N)$ and grouped them into the $\nu N$-dimensional vector
$\vec{\bf p}(\vec\sigma) = 
({\bf p}_1(\sigma_1),{\bf p}_2(\sigma_2),\ldots,{\bf p}_N(\sigma_N))$, 
which couples via the scalar product to the $\vec{\bf r}$-vector in 
the exponential of equation (\ref{3.6}).
Since the momentum vector $\vec{\bf p}(\vec\sigma)$ satisfies the requirement 
$\vert \vec{\bf p}(\vec\sigma)\vert = \vert \vec{\bf p}\vert$
for an arbitrary state vector $\vec\sigma$, one can apply the key
relation (\ref{3.5}) to every summand in the exponential representation
(\ref{3.6}) of $\psi_{\vec{\bf p}}(\vec{\bf r})$ to obtain
\begin{eqnarray} 
\psi^*_{\vec{\bf p}}(\vec{{\bf r}})
\frac{1}{H+z} \psi_{\vec{\bf p}}(\vec{{\bf r}}) & = &
\psi^*_{\vec{\bf p}}(\vec{{\bf r}})
\frac{1}{(\sqrt{2}{\rm i})^N} \sum_{\{ \vec{\sigma}\}} 
\left( \prod_{j=1}^N \sigma_j \right) 
{\rm e}^{({\rm i}/\hbar)\vec{\bf p}(\vec{\sigma})\cdot\vec{\bf r}} 
\nonumber \\ & & \times
\frac{1}{D} \sum_{n=0}^{\infty} \left\{
\left[ \frac{{\rm i}\hbar}{m} \vec{\bf p}(\vec{\sigma}) \cdot \bfnabla
+ \frac{\hbar^2}{2 m} \bfnabla^2 \right] \frac{1}{D} \right\}^n . 
\label{3.7} 
\end{eqnarray}

Let us consider the term of order $n$ in the expansion (\ref{3.7})
\begin{equation} \label{3.8} 
\fl \prod_{j=1}^N \left[ \sin\left( \frac{p_j^x x_j}{\hbar} \right) 
\right] \frac{1}{{\rm i}^N} \sum_{\{ \vec{\sigma}\}} 
\prod_{j=1}^N \left( \sigma_j {\rm e}^{({\rm i}/\hbar)\sigma_j p_j^x x_j} 
\right) \frac{1}{D} \left\{ \left[ 
\frac{{\rm i}\hbar}{m} \vec{\bf p}(\vec{\sigma}) \cdot \bfnabla
+ \frac{\hbar^2}{2 m} \bfnabla^2 \right] \frac{1}{D} \right\}^n . 
\end{equation}
One first carries out all differentiations in (\ref{3.8}).
Since the scalar product
\begin{equation} \label{3.9}
\vec{\bf p}(\vec{\sigma}) \cdot \bfnabla = 
\sum_{k=1}^N \sigma_k p_k^x \frac{\partial}{\partial x_k} +
\sum_{k=1}^N {\bf p}_k^{\perp}\cdot \enabla_{{\bf r}_k^{\perp}}
\end{equation}
contains components of the state vector $\vec{\sigma}$, each of 
the components can take either odd or even power multiplied by
the corresponding exponential; the summation over $\sigma_j=\pm 1$
then gives rise to the original stationary wave
$\propto \sin(p_j^x x_j/\hbar)$ for odd powers of $\sigma_j$
and to the shifted wave $\propto \cos(p_j^x x_j/\hbar)$ for
even powers of $\sigma_j$.
As the next step, one performs the inverse Laplace transform of
$1/D^j$ $(j=1,2,\ldots)$ using the formula (\ref{2.11}).
In the consequent relatively simple integrations over momenta 
we have to keep in mind that while the integration of 
odd powers of the ${\bf p}^{\perp}$ components give zero contributions,
this is no longer true for odd powers of $p^x$ since the integration
goes only over the positive values of $p^x$.
The final result for the Boltzmann density in configuration space
(\ref{3.1}) has the form analogous to that of the bulk case (\ref{2.12}),
$\langle \vec{{\bf r}} \vert {\rm e}^{-\beta H} \vert \vec{{\bf r}} \rangle
= \sum_{n=0}^{\infty} B_{\beta}^{(n)}(\vec{\bf r})$
where $B_{\beta}^{(n)}(\vec{\bf r})$ denotes the contribution
implied by the $n$th term (\ref{3.8}). 

We have performed the outlined procedure for the first terms
$B_{\beta}^{(0)}$, $B_{\beta}^{(1)}$ and $B_{\beta}^{(2)}$.
We do not go into details of the calculations, but only write down
the final result and then discuss the origin of the terms obtained 
and their structure.

The result for $B_{\beta}^{(0)}(\vec{\bf r})$ reads
\begin{eqnarray} 
B_{\beta}^{(0)}(\vec{\bf r}) & = &
\int_0^{\infty} \frac{{\rm d}\vec{p}^x}{(\pi\hbar)^N}
\int \frac{{\rm d}\vec{\bf p}^{\perp}}{(2\pi\hbar)^{(\nu-1)N}} 
\prod_{j=1}^N \left[ 2 \sin^2\left( \frac{p_j^x x_j}{\hbar} \right) \right] 
{\rm e}^{-\beta[\vec{p}^2/(2m) + V(\vec{r})]} \nonumber \\
& = & \frac{1}{(\sqrt{2\pi} \lambda)^{\nu N}} {\rm e}^{-\beta V} 
\prod_{j=1}^N \left( 1 - {\rm e}^{-2x_j^2/\lambda^2} \right) .  \label{3.10}
\end{eqnarray}
In comparison with the corresponding bulk term (\ref{2.13a}),
each particle gets an additional ``boundary'' factor
$1-\exp(-2 x^2/\lambda^2)$ which goes from $0$ at the boundary $x=0$
to $1$ in the bulk interior $x\to\infty$ on the length scale $\sim\lambda$.
The product of boundary factors ensures that the quantum Boltzmann
density vanishes as soon as one of the particles lies on the boundary.
The dependence of the boundary factor on the de Broglie wavelength
$\lambda$ is non-analytic; this fact prevents one from a simple classification
of contributions to the Boltzmann density according to integer powers 
of $\lambda$ like it was in the bulk case.
However, as will be shown explicitly in the next section on the evaluation of 
the partition function, when the exponential part $\exp(-2 x^2/\lambda^2)$ 
of the boundary factor (multiplied eventually by another 
$\lambda$-independent function of $x$) is integrated over the $x$-coordinate,
the analyticity in the parameter $\lambda$ is restored.
At this stage we only notice that the integrated product of boundary 
factors should be expanded in the following way
\begin{equation} \label{3.11}
\prod_{j=1}^N \left( 1 - {\rm e}^{-2x_j^2/\lambda^2} \right)
= 1 - \sum_{j=1}^N {\rm e}^{-2x_j^2/\lambda^2} 
+ \frac{1}{2!} \sum_{j,k=1\atop (j\ne k)}^N {\rm e}^{-2x_j^2/\lambda^2} 
{\rm e}^{-2x_k^2/\lambda^2} + \cdots , 
\end{equation} 
where each exponential term $\exp(-2 x^2/\lambda^2)$, when integrated
over $x$, produces one $\lambda$-factor as the result of 
the evoqued substitution of variables $x=\lambda x'$.
 
The result for $B_{\beta}^{(1)}(\vec{\bf r})$ is found in the form
\begin{eqnarray}
B_{\beta}^{(1)}(\vec{\bf r}) & = & 
\frac{1}{(\sqrt{2\pi} \lambda)^{\nu N}} {\rm e}^{-\beta V} \Bigg\{ 
- \sum_{k=1}^N 
\prod_{j=1\atop (j\ne k)}^N \left( 1 - {\rm e}^{-2x_j^2/\lambda^2} \right) 
x_k {\rm e}^{-2x_k^2/\lambda^2}
\beta \frac{\partial V}{\partial x_k} \nonumber \\ 
& & + \prod_{j=1}^N \left( 1 - {\rm e}^{-2x_j^2/\lambda^2} \right)
\lambda^2 \left[ - \frac{\beta}{4} \bfnabla^2 V +
\frac{\beta^2}{6} \left( \bfnabla V \right)^2 \right] \Bigg\} . \label{3.12}
\end{eqnarray}
Here, the first term corresponds to the source operator 
$D^{-1} \vec{\bf p}(\vec\sigma)\cdot \bfnabla D^{-1}$ whose bulk 
counterpart does not contribute to the Boltzmann density (\ref{2.13b}).
The dependence on $\lambda$ appears, besides the product of 
the particle boundary factors $1-\exp(-2x^2/\lambda^2)$, also via
the combination $x \exp(-2x^2/\lambda^2)$.
This function has a maximum of order $\lambda$ and the integration of 
this function over $x$ gives a contribution of order $\lambda^2$.
These properties make the function $x \exp(-2x^2/\lambda^2)$
of an order equal to or ``weaker'' than $\lambda$, and therefore this
function is a legitimate expansion parameter.
The second term on the rhs of equation (\ref{3.12}), having the origin in 
the source operator $D^{-1}\bfnabla^2D^{-1}$, corresponds to the simple
multiplication of its bulk counterpart (\ref{2.13b}) by the product of 
particle boundary factors $1-\exp(-2x^2/\lambda^2)$.

The final formula for $B_{\beta}^{(2)}(\vec{\bf r})$ consists in four terms
\begin{equation}
B_{\beta}^{(2)}(\vec{\bf r}) = B_{\beta}^{(2,1)}(\vec{\bf r}) +
B_{\beta}^{(2.2)}(\vec{\bf r}) + B_{\beta}^{(2,3)}(\vec{\bf r}) +
B_{\beta}^{(2,4)}(\vec{\bf r}) , \label{3.13}
\end{equation}
where
\numparts
\begin{eqnarray}
\fl B_{\beta}^{(2,1)}(\vec{\bf r}) & = &
\frac{1}{(\sqrt{2\pi} \lambda)^{\nu N}} {\rm e}^{-\beta V} \Bigg\{
\prod_{j=1}^N \left( 1 - {\rm e}^{-2x_j^2/\lambda^2} \right)
\lambda^2 \left[ \frac{\beta}{6} \bfnabla^2 V 
- \frac{\beta^2}{8} \left( \bfnabla V\right)^2 \right] \nonumber \\ 
\fl & & + \sum_{k=1}^N 
\prod_{j=1\atop (j\ne k)}^N \left( 1 - {\rm e}^{-2x_j^2/\lambda^2} \right) 
x_k^2 {\rm e}^{-2x_k^2/\lambda^2}
\left[ \frac{2\beta}{3} \frac{\partial^2 V}{\partial x_k^2} -
\frac{\beta^2}{2} \left( \frac{\partial V}{\partial x_k} \right)^2 \right] 
\label{3.14a} \\ \fl & & - \sum_{k,l=1\atop (k\ne l)}^N 
\prod_{j=1\atop (j\ne k,l)}^N \left( 1 - {\rm e}^{-2x_j^2/\lambda^2} \right) 
x_k {\rm e}^{-2x_k^2/\lambda^2} x_l {\rm e}^{-2x_l^2/\lambda^2}
\left[ \frac{2\beta}{3} \frac{\partial^2 V}{\partial x_k \partial x_l} -
\frac{\beta^2}{2} \frac{\partial V}{\partial x_k} 
\frac{\partial V}{\partial x_l} \right] \Bigg\} \nonumber
\end{eqnarray}
corresponds to the source operator
$D^{-1} \vec{\bf p}(\vec\sigma)\cdot \bfnabla [ D^{-1}
\vec{\bf p}(\vec\sigma)\cdot \bfnabla D^{-1}]$,
\begin{eqnarray}
\fl B_{\beta}^{(2,2)}(\vec{\bf r}) & = &
\frac{1}{(\sqrt{2\pi} \lambda)^{\nu N}} {\rm e}^{-\beta V}
\sum_{k=1}^N \prod_{j=1\atop (j\ne k)}^N 
\left( 1 - {\rm e}^{-2x_j^2/\lambda^2} \right) 
x_k {\rm e}^{-2x_k^2/\lambda^2} \lambda^2 \left[ 
- \frac{\beta}{6} \frac{\partial}{\partial x_k}
\left( \bfnabla^2 V \right) \right. \nonumber \\
\fl & & \left. +
\frac{\beta^2}{8} \left( \bfnabla^2 V \right) \frac{\partial V}{\partial x_k}
+ \frac{\beta^2}{12} \frac{\partial}{\partial x_k} \left( \bfnabla V \right)^2
-\frac{\beta^3}{15} \left( \bfnabla V \right)^2 
\frac{\partial V}{\partial x_k} \right] \label{3.14b} 
\end{eqnarray}
has its origin in the source operator
$D^{-1} \vec{\bf p}(\vec\sigma)\cdot \bfnabla [ D^{-1}
\bfnabla^2 D^{-1}]$,
\begin{eqnarray}
\fl B_{\beta}^{(2,3)}(\vec{\bf r}) & = &
\frac{1}{(\sqrt{2\pi} \lambda)^{\nu N}} {\rm e}^{-\beta V}
\sum_{k=1}^N \prod_{j=1\atop (j\ne k)}^N 
\left( 1 - {\rm e}^{-2x_j^2/\lambda^2} \right) 
x_k {\rm e}^{-2x_k^2/\lambda^2} \lambda^2 \left[ 
- \frac{\beta}{6} \frac{\partial}{\partial x_k}
\left( \bfnabla^2 V \right) \right. \nonumber \\
\fl & & \left. +
\frac{\beta^2}{8} \left( \bfnabla^2 V \right) \frac{\partial V}{\partial x_k}
+ \frac{\beta^2}{8} \frac{\partial}{\partial x_k} \left( \bfnabla V \right)^2
-\frac{\beta^3}{10} \left( \bfnabla V \right)^2 
\frac{\partial V}{\partial x_k} \right] \label{3.14c} 
\end{eqnarray}
comes from the source operator
$D^{-1} \bfnabla^2 [ D^{-1}
\vec{\bf p}(\vec\sigma)\cdot \bfnabla D^{-1}]$, and
\begin{eqnarray}
\fl B_{\beta}^{(2,4)}(\vec{\bf r}) & = &
\frac{1}{(\sqrt{2\pi} \lambda)^{\nu N}} {\rm e}^{-\beta V}
\prod_{j=1}^N \left( 1 - {\rm e}^{-2x_j^2/\lambda^2} \right) 
\lambda^4 \left[ -\frac{\beta}{24} \left(\bfnabla^2 \right)^2 V
\right. \nonumber \\ 
\fl & & \left. 
+\frac{\beta^2}{16} \bfnabla V \cdot \bfnabla\left( \bfnabla^2 V\right)
+\frac{\beta^2}{48} \bfnabla^2\left( \bfnabla V\right)^2 
+\frac{\beta^2}{32} \left(\bfnabla^2 V\right)^2 
\right. \label{3.14d} \\ 
\fl & & \left.
-\frac{\beta^3}{30} \bfnabla V \cdot \bfnabla\left( \bfnabla V\right)^2
-\frac{\beta^3}{24} \left(\bfnabla V\right)^2 \bfnabla^2 V
+ \frac{\beta^4}{72}\left(\bfnabla V\right)^4 \right] \nonumber
\end{eqnarray}
\endnumparts
results from the source operator $D^{-1}\bfnabla^2[D^{-1}\bfnabla^2 D^{-1}]$.
It is easy to check that the term $B_{\beta}^{(2)}(\vec{\bf r})$
as a whole is of the leading order $\lambda^2$, like it was in 
the bulk regime (see equation (\ref{2.13c})). 

We would like to note that in the bulk interior far away from the boundary, 
i.e., when all $x$-coordinates of particles $\{ x_j\to\infty \}_{j=1}^N$, 
the results for the boundary Boltzmann density must reduce to the bulk ones 
(\ref{2.13a})--(\ref{2.13c}).
It is easy to verify that the results obtained pass this test.

To summarize, the $\lambda$-expansion of the Boltzmann density in the
presence of a boundary is more complex than the analytic one in the bulk. 
The expansion involves not only powers of $\lambda$, but also
non-analytic position-dependent terms of type
$1-\exp(-2 x^2/\lambda^2)$, $x \exp(-2 x^2/\lambda^2)$, etc. 
The first three Boltzmann terms $B_{\beta}^{(0)}(\vec{\bf r})$,   
$B_{\beta}^{(1)}(\vec{\bf r})$ and $B_{\beta}^{(2)}(\vec{\bf r})$ 
exhibit properties similar to their bulk counterparts
(\ref{2.13a})--(\ref{2.13b}): The maximum of 
$B_{\beta}^{(n)}(\vec{\bf r})$ is of order $\lambda^n$.
We anticipate that this formal structure of the Boltzmann terms
is also maintained on higher levels.  

\section{Half-space geometry: statistical quantities}

\subsection{Partition function and free energy}
Since the Boltzmann density is represented by the series
$\sum_{n=0}^{\infty} B_{\beta}^{(n)}(\vec{\bf r})$,
the partition function (\ref{2.15}) is expressible as
\begin{equation} \label{4.1}
Z_{\rm qu} = \sum_{n=0}^{\infty} Z_{\rm qu}^{(n)} , \quad
Z_{\rm qu}^{(n)} = \frac{1}{N!} \int_{\Lambda} {\rm d}\vec{\bf r}\,
B_{\beta}^{(n)}(\vec{\bf r}) .    
\end{equation}
In the next paragraphs we evaluate consecutively the $\lambda$-expansion
of the first three terms $Z_{\rm qu}^{(0)}$, $Z_{\rm qu}^{(1)}$ and 
$Z_{\rm qu}^{(2)}$, up to the order $\lambda^2$.
According to the analysis in the previous section, higher-order
$Z_{\rm qu}^{(n)}$ terms are expected to contribute only to $\lambda^3$
and higher powers of $\lambda$.

Inserting the Boltzmann term $B_{\beta}^{(0)}$ (\ref{3.10}) into 
the definition (\ref{4.1}) of $Z_{\rm qu}^{(0)}$ and performing 
the expansion (\ref{3.11}) for the product of boundary factors, we obtain
\begin{equation} \label{4.2}
\fl \frac{Z_{\rm qu}^{(0)}}{Z} = 1
- \int_{\Lambda} {\rm d}{\bf r}\, {\rm e}^{-2x^2/\lambda^2} n({\bf r})
+ \frac{1}{2!} \int_{\Lambda} {\rm d}{\bf r}_1 
\int_{\Lambda} {\rm d}{\bf r}_2\, {\rm e}^{-2x_1^2/\lambda^2}
{\rm e}^{-2x_2^2/\lambda^2} n^{(2)}({\bf r}_1,{\bf r}_2) + \cdots .
\end{equation}
The system is translationally invariant in the ${\bf r}^{\perp}$ space,
i.e. $n({\bf r}) = n(x)$, $n^{(2)}({\bf r}_1,{\bf r}_2) = 
n^{(2)}(x_1,x_2;\vert {\bf r}^{\perp}_1-{\bf r}^{\perp}_2\vert)$, etc.
This property enables us to rewrite equation (\ref{4.2}) in a more convenient
form 
\begin{eqnarray} 
\fl \frac{Z_{\rm qu}^{(0)}}{Z} & = & 1 - \vert\partial\Lambda\vert
\int_0^{\infty} {\rm d}x\, {\rm e}^{-2x^2/\lambda^2} n(x)
+ \frac{1}{2!} \vert\partial\Lambda\vert^2
\left[ \int_0^{\infty} {\rm d}x\, {\rm e}^{-2x^2/\lambda^2} n(x)\right]^2
\nonumber \\ \fl & &
+ \frac{1}{2!} \vert\partial\Lambda\vert \int {\rm d}{\bf r}^{\perp} 
\int_0^{\infty} {\rm d}x_1 \int_0^{\infty} {\rm d}x_2\, 
{\rm e}^{-2x_1^2/\lambda^2} {\rm e}^{-2x_2^2/\lambda^2} 
n^{(2){\rm T}}(x_1,x_2;\vert {\bf r}^{\perp}\vert ) + \cdots , \label{4.3}
\end{eqnarray}
where $\vert\partial\Lambda\vert$ denotes the surface of the half-space 
domain $\Lambda$ at $x=0$, and the third term has been added to the rhs and 
subsequently subtracted from the double integral (inducing in this way
the truncation of the two-body density) in order to arrange the series 
for the application of the cumulant method. 
Assuming that the classical density profile $n(x)$ is analytic
at the boundary $x=0$, i.e. $n(x) = n(0) +  n'(0) x + 
n''(0) x^2/2! + \cdots$ (this condition is not always fulfilled, see 
subsection 5.1.), the integral
\begin{eqnarray} 
\int_0^{\infty} {\rm d}x\, {\rm e}^{-2x^2/\lambda^2} n(x)
& = & \lambda \int_0^{\infty} {\rm d}x'\, {\rm e}^{-2x'^2} n(\lambda x')
\nonumber \\ & = & \lambda \frac{1}{2} \sqrt{\frac{\pi}{2}} n(0)
+ \lambda^2 \frac{1}{4} n'(0) + O(\lambda^3) . \label{4.4}
\end{eqnarray}
Performing an analogous procedure in the double integral on the rhs of
equation (\ref{4.3}), we finally arrive at
\begin{eqnarray} 
\frac{Z_{\rm qu}^{(0)}}{Z} & = & 1 - \vert\partial\Lambda\vert
\lambda \frac{1}{2} \sqrt{\frac{\pi}{2}} n(0) + \frac{1}{2!}
\left[ \vert\partial\Lambda\vert \lambda \frac{1}{2} 
\sqrt{\frac{\pi}{2}} n(0) \right]^2 
- \vert\partial\Lambda\vert \lambda^2 \frac{1}{4} n'(0) \nonumber \\
& & + \vert\partial\Lambda\vert \lambda^2 \frac{\pi}{16}
\int {\rm d}{\bf r}^{\perp}\, 
n^{(2){\rm T}}(0,0;\vert {\bf r}^{\perp}\vert ) + O(\lambda^3) .\label{4.5} 
\end{eqnarray}
We recall that the boundary values of the classical statistical quantities
$n(0)$, $n'(0)$ and $n^{(2){\rm T}}(0,0;\vert {\bf r}^{\perp}\vert)$
are, in general, nonzero.

The boundary factors $1-\exp(-2x^2/\lambda^2)$ do not play any role
in the Boltzmann term $B_{\beta}^{(1)}$ (\ref{3.12}) when one is 
interested in the $\lambda$-expansion of the corresponding $Z_{\rm qu}^{(1)}$
up to the $\lambda^2$ order, so we can ignore them.
After simple algebra, $Z_{\rm qu}^{(1)}$ is written as
\begin{equation} \label{4.6}
\fl \frac{Z_{\rm qu}^{(1)}}{Z} = \int_{\Lambda} {\rm d}{\bf r}\, 
{\rm e}^{-2x^2/\lambda^2} x n'(x) + 
\lambda^2 \left[ - \frac{\beta}{4} \left\langle \bfnabla^2 V \right\rangle 
+ \frac{\beta^2}{6} \left\langle (\bfnabla V)^2 \right\rangle \right] 
+ O(\lambda^3) , 
\end{equation}
where the classical average is taken over the half-space $\Lambda$.
The integral on the rhs of (\ref{4.6}) can be treated in the way
outlined in the previous paragraph, with the result
\begin{equation} \label{4.7}
\int_{\Lambda} {\rm d}{\bf r}\, {\rm e}^{-2x^2/\lambda^2} x n'(x) 
= \vert\partial\Lambda\vert \lambda^2 \frac{1}{4} n'(0) + O(\lambda^3) ,  
\end{equation} 
the second term can be simplified by using the evident equality
\begin{eqnarray}
\beta^2 \left\langle (\bfnabla V)^2 \right\rangle & = &
\beta \left\langle \bfnabla^2 V \right\rangle 
+ \int_{\Lambda} {\rm d}{\bf r}\, \enabla^2 n({\bf r}) \nonumber \\
& = & \beta \left\langle \bfnabla^2 V \right\rangle 
- \vert\partial\Lambda\vert n'(0) , \label{4.8}
\end{eqnarray}
provided that $n'(0)$ is finite, see subsection 5.1.

 From the four contributions to the Boltzmann term 
$B_{\beta}^{(2)}(\vec{\bf r})$ in (\ref{3.13}), only the first one
implies the $\lambda^2$ term:
\begin{equation} \label{4.9}
\frac{Z_{\rm qu}^{(2)}}{Z} = \lambda^2 \left[ \frac{\beta}{6} 
\left\langle \bfnabla^2 V \right\rangle - \frac{\beta^2}{8} 
\left\langle (\bfnabla V)^2 \right\rangle \right] 
+ O(\lambda^3) .
\end{equation}
Also here we can apply the relation (\ref{4.8}) to eliminate
the term $\beta^2\langle(\bfnabla V)^2\rangle$ in favour of 
the Laplacian term.

The results obtained in this subsection can be summarized by the expansion
formula for the free energy, expanding in powers of $\lambda$ 
the logarithm in $\beta F_{\rm qu} = - \ln Z_{\rm qu}$,
\begin{eqnarray} 
\beta F_{\rm qu} & = &\beta F + \lambda^2 \frac{\beta}{24}
\left\langle \bfnabla^2 V \right\rangle 
+ \vert\partial\Lambda\vert \left\{ 
\lambda \frac{1}{2} \sqrt{\frac{\pi}{2}} n(0)
\right. \nonumber \\ & & \left.
+ \lambda^2 \frac{1}{24} n'(0)
- \lambda^2 \frac{\pi}{16} \int {\rm d}{\bf r}^{\perp}\, 
n^{(2){\rm T}}(0,0;\vert {\bf r}^{\perp}\vert ) \right\}
+ O(\lambda^3) . \label{4.10}
\end{eqnarray}
The free energy of a quantum particle system constrained to
a domain $\Lambda$ of volume $\vert\Lambda\vert$ and 
surface $\vert\partial\Lambda\vert$ possesses the following
general form
\begin{equation} \label{4.11}
F_{\rm qu} = f_{\rm qu}^{b} \vert\Lambda\vert
+\gamma_{\rm qu} \vert\partial\Lambda\vert ,
\end{equation}  
where $f_{\rm qu}^{b}$ is the bulk free energy per unit volume and
$\gamma_{\rm qu}$ is the surface tension.
In the formula (\ref{4.10}), the quantum contribution to the
bulk part of the free energy comes only from the potential-dependent
term $\propto \langle \bfnabla^2 V \rangle$.
This term can, in principle, give also some surface contribution.
But for our model system of the semi-infinite one-component plasma constrained 
by the plain hard wall, the potential $V$ fulfils the Poisson equation 
(\ref{2.23}) and therefore the term $\propto \langle \bfnabla^2 V \rangle$ 
produces only the bulk quantum contribution to the classical free energy, 
as shown in equation (\ref{2.25}).
By comparing (\ref{4.10}) with (\ref{4.11}), the surface tension
is found to be
\begin{equation} \label{4.12}
\fl \beta \gamma_{\rm qu} = \beta \gamma
+ \lambda \frac{1}{2} \sqrt{\frac{\pi}{2}} n(0)
+ \lambda^2 \frac{1}{24} n'(0)
- \lambda^2 \frac{\pi}{16} \int {\rm d}{\bf r}^{\perp}\, 
n^{(2){\rm T}}(0,0;\vert {\bf r}^{\perp}\vert ) + O(\lambda^3) .
\end{equation}
It is interesting that up to the $\lambda^2$ order the interaction 
potential $V$ enters into the formula for the surface tension only 
implicitly via the classical one-body and two-body averages.
We note that, although for the one-component plasma the boundary truncated 
two-body density $n^{(2){\rm T}}(0,0;\vert {\bf r}^{\perp}\vert )$ in 
(\ref{4.12}) exhibits a long-range decay of type 
$1/\vert {\bf r}^{\perp}\vert^{\nu}$ at asymptotically large 
$\vert {\bf r}^{\perp}\vert\to\infty$ \cite{Jancovici82a,Jancovici82b}, 
this makes no divergence problem at large distances in the integration 
over the $(\nu-1)$-dimensional ${\bf r}^{\perp}$ space.  

Without going into details, the result for the surface tension (\ref{4.12})
can be straightforwardly generalized to many-component quantum fluids
composed of different species $\alpha = 1,2,\ldots,L$ with masses
$\{ m_{\alpha} \}_{\alpha=1}^L$ and the corresponding de Broglie wavelengths  
$\lambda_{\alpha}=\hbar(\beta/m_{\alpha})^{1/2}$:
\begin{eqnarray} 
\beta \gamma_{\rm qu} & = & \beta \gamma
+ \frac{1}{2} \sqrt{\frac{\pi}{2}} \sum_{\alpha} \lambda_{\alpha} n_{\alpha}(0)
+ \frac{1}{24} \sum_{\alpha} \lambda_{\alpha}^2 n_{\alpha}'(0)
\nonumber \\ & &
- \frac{\pi}{16} \int {\rm d}{\bf r}^{\perp}\, \sum_{\alpha,\beta}
\lambda_{\alpha} \lambda_{\beta}
n^{(2){\rm T}}_{\alpha\beta}(0,0;\vert {\bf r}^{\perp}\vert ) 
+ O(\lambda_{\alpha} \lambda_{\beta} \lambda_{\gamma}) . \label{4.13}
\end{eqnarray}
We note that in the special case of a multi-component Coulomb fluid containing 
at least two species with charges of opposite signs, a short-distance
regularization of the Coulomb potential is needed in the evaluation of 
the classical averages in (\ref{4.13}) in order to prevent the thermodynamic
collapse. 

\subsection{Particle density profile}
As concerns the quantum corrections to the classical density profile
and two-body density, since they involve a quantum Boltzmann factor which is
not integrated on all variables, their expansion will contain, in addition to
powers of $\lambda$, non-analytic terms $\exp(-2x^2/\lambda^2)$. Here, we
shall restrict ourselves to the leading correction of order $\lambda$,
considering a factor $x\exp(-2x^2/\lambda^2)$ as being of order $\lambda$
since its maximum is of order $\lambda$, as pointed out in section 3, using
the expansion technique presented in the previous subsection 4.1.  

The one-body density profile (\ref{2.17}) is obtained in the form
\begin{eqnarray} 
n_{\rm qu}(x) & = & \left( 1 - {\rm e}^{-2x^2/\lambda^2} \right)
\left[ n(x) - \lambda \frac{1}{2} \sqrt{\frac{\pi}{2}}
\int {\rm d}{\bf r}^{\perp}\, n^{(2){\rm T}}(x,0;\vert{\bf r}^{\perp}\vert)
\right] \nonumber \\
& & + x {\rm e}^{-2x^2/\lambda^2} n'(x) + O(\lambda^2) , \label{4.14}
\end{eqnarray}
where the first (product) term has its origin in the Boltzmann term
$B_{\beta}^{(0)}(\vec{\bf r})$ (\ref{3.10}) and the second term comes
from $B_{\beta}^{(1)}(\vec{\bf r})$ (\ref{3.12}).

In the classical description of the one-component Coulomb plasma,
the system is neutral as a whole,
\begin{equation} \label{4.15}
\int_0^{\infty} {\rm d}x\, \left[ n(x) - n \right] = 0 ,
\end{equation}
and a particle of charge $e$ is surrounded by a screening cloud
the average charge of which is exactly $-e$,
\begin{equation} \label{4.16}
n(x) = - \int {\rm d}{\bf r}^{\perp} \int_0^{\infty} {\rm d}x'\,
n^{(2){\rm T}}(x,x';\vert{\bf r}^{\perp}\vert) .
\end{equation} 
The neutrality sum rule (\ref{4.15}) is valid also in the quantum regime.
For our purposes it can be reexpressed in the form
\begin{equation} \label{4.17}
\int_0^{\infty} {\rm d}x \left[ n_{\rm qu}(x) - n(x) \right] = 0 .
\end{equation}
It is easy to check that for the density profile (\ref{4.14})
this relation is fulfilled in the linear $\lambda$ order.
 
The formula (\ref{4.14}) can be straightforwardly extended to
a quantum fluid with many components $\alpha=1,2,\ldots,L$:
\begin{eqnarray} 
n_{\rm qu}^{(\alpha)}(x) & = & \left( 1 - {\rm e}^{-2x^2/\lambda_{\alpha}^2} 
\right) \left[ n_{\alpha}(x) - \frac{1}{2} \sqrt{\frac{\pi}{2}}
\int {\rm d}{\bf r}^{\perp}\, \sum_{\beta} \lambda_{\beta} 
n^{(2){\rm T}}_{\alpha\beta}(x,0;\vert{\bf r}^{\perp}\vert)
\right] \nonumber \\
& & + x {\rm e}^{-2x^2/\lambda_{\alpha}^2} n'_{\alpha}(x) + 
O(\lambda^2) . \label{4.18}
\end{eqnarray}

\subsection{Two-body density}
The leading quantum correction for the truncated two-body density (\ref{2.19})
is found to be
\numparts
\begin{eqnarray}
n_{\rm qu}^{(2){\rm T}}({\bf r}_1,{\bf r}_2) & = & 
\left( 1 - {\rm e}^{-2x_1^2/\lambda^2} \right)
\left( 1 - {\rm e}^{-2x_2^2/\lambda^2} \right)
\Bigg\{ n^{(2){\rm T}}({\bf r}_1,{\bf r}_2)  
\nonumber \\ & &
- \lambda \frac{1}{2} \sqrt{\frac{\pi}{2}} \int {\rm d}{\bf r}^{\perp}\, 
n^{(3){\rm T}}[{\bf r}_1, {\bf r}_2, (0,{\bf r}^{\perp})]
\Bigg\} \nonumber \\
& & + \left( 1 - {\rm e}^{-2x_2^2/\lambda^2} \right) 
x_1 {\rm e}^{-2x_1^2/\lambda^2} \frac{\partial}{\partial x_1}
n^{(2){\rm T}}({\bf r}_1,{\bf r}_2)   \label{4.19a} \\
& & + \left( 1 - {\rm e}^{-2x_1^2/\lambda^2} \right) 
x_2 {\rm e}^{-2x_2^2/\lambda^2} \frac{\partial}{\partial x_2}
n^{(2){\rm T}}({\bf r}_1,{\bf r}_2) + O(\lambda^2) , \nonumber
\end{eqnarray}
where
\begin{eqnarray}
\fl n^{(3){\rm T}}\left( {\bf r}_1, {\bf r}_2, {\bf r}_3 \right)
& = & n^{(3)}\left( {\bf r}_1, {\bf r}_2, {\bf r}_3 \right)
- n^{(2){\rm T}}\left( {\bf r}_1, {\bf r}_2 \right) n({\bf r}_3)
- n^{(2){\rm T}}\left( {\bf r}_1, {\bf r}_3 \right) n({\bf r}_2)
\nonumber \\ \fl & & 
- n^{(2){\rm T}}\left( {\bf r}_2, {\bf r}_3 \right) n({\bf r}_1)
- n({\bf r}_1) n({\bf r}_2) n({\bf r}_3) . \label{4.19b}
\end{eqnarray}
\endnumparts

\section{Coulomb models with known classical statistical quantities}

\subsection{High-temperature Debye-H\"{u}ckel limit}
For the semi-infinite one-component plasma, at least in two and three
dimensions, the classical truncated two-body density and density profile are
known in this limit (see, e.g., equations (3.12) and (3.27) in 
\cite{Jancovici82a}). 
The classical surface tension is also known (see equations (4.6) and (4.7) in 
\cite{Russier}). 
The integrals which appear in (\ref{4.12}) and (\ref{4.14}) 
are easily computed:
\begin{equation} \label{5.1}
\int{\rm d}{\bf r}^{\perp}n^{(2){\rm T}}(x,0;|{\bf r}^{\perp}|)
=-n\kappa {\rm e}^{-\kappa x}, 
\end{equation}
where $\kappa=[2(\nu-1)\pi\beta e^2 n]^{1/2}$ is the inverse Debye length.
It should be noted that the density profile
\begin{eqnarray}
n(x) & = & n+\frac{\kappa^{\nu}}{2\pi(\nu-1)} \nonumber \\
& &\times\int_0^{\infty}\frac{{\rm e}^{-\kappa x}-2(1+t^2)^{1/2}
\exp[-2(1+t^2)^{1/2}\kappa x]}{[(1+t^2)^{1/2}+t]^2(3+4t^2)}t^{\nu-2}{\rm d}t,
\label{5.2}
\end{eqnarray}
contrarily to the assumption in subsection 4.1., can be expanded in
powers of $x$ only up to order $x^{3-\nu}$: the $n$th derivative of $n(x)$ is
infinite for $n\geq 4-\nu$. 

In two dimensions, $n'(0)$ is finite and the quantum surface tension up to
order $\lambda^2$ (\ref{4.12}) becomes 
\begin{eqnarray}
\beta \gamma_{\rm qu} & = & \frac{\kappa}{2\pi}\left(1-\frac{\pi}{4}\right) 
+ \lambda \frac{1}{2} \sqrt{\frac{\pi}{2}}
\left[n-\left(\ln 3-1+\frac{\pi\sqrt{3}}{9}\right)\frac{\kappa^2}{8\pi}\right]
\nonumber \\ 
& & + \lambda^2 \left(\frac{\kappa^3}{72\pi}
+ \frac{\pi}{16}n\kappa \right)+ o(\lambda^2) . \label{5.3}
\end{eqnarray}
It is likely that $n''(0)$, which is infinite, would appear if we attempted to
compute the term of order $\lambda^3$. 
This is an indication that the Wigner-Kirkwood expansion of the surface 
tension is \emph{not} in integer powers of $\lambda$ in the present case; 
some singular term (logarithmic for instance) appears beyond the order 
$\lambda^2$.
 
In three dimensions, $n'(0)$ is infinite. 
The integral (\ref{4.4}) is nevertheless finite, although it cannot 
be expanded up to order $\lambda^2$ (instead, a term of order 
$\lambda^2\ln\lambda$ appears).
Also, using (\ref{4.8}) in the contributions (\ref{4.6}) and (\ref{4.9}) to
the partition function shows that these contributions, formally of order
$\lambda^2$, are infinite. 
This is an indication that some singular term, which we are unable 
to estimate, appears in the Wigner-Kirkwood expansion of
the surface tension beyond the order $\lambda$. 
At order $\lambda$, (\ref{4.12}) gives 
\begin{equation} \label{5.4}
\beta \gamma_{\rm qu} = \frac{2\ln2-1}{32\pi}\kappa^2
+\lambda\frac{1}{2}\sqrt{\frac{\pi}{2}}\left[n-\left(\frac{1-3\ln3+\pi\sqrt{3}}
{4}\frac{\kappa^3}{24\pi}\right)\right]+o(\lambda) . 
\end{equation}

For the two-component plasma, the classical density profiles (see equation 
(4.9) in \cite{Jancovici82a}), which contain the singular Bessel function
$K_2(2\kappa x)$, also cannot be expanded in powers of $x$ to all orders.

It is also interesting to compare the density profiles (\ref{4.18}) for a
many-component semi-infinite fluid with the results of Aqua and Cornu 
\cite{Aqua04}, which have been obtained by a very different method, for a
three-dimensional multi-component semi-infinite quantum Coulomb fluid. 
Let the charge of a particle of species $\alpha$ be $e_{\alpha}$. 
The inverse Debye length now is 
$\kappa=(4\pi\beta\sum_{\alpha}n_{\alpha}e_{\alpha}^2)^{1/2}$. 
An easy generalization of (\ref{5.1}) shows that 
\begin{equation} \label{5.5}
\int{\rm d}{\bf r}^{\perp}n_{\alpha\beta}^{(2){\rm T}}(x,0;|{\bf r}^{\perp}|)
=-\frac{4\pi\beta n_{\alpha}e_{\alpha}n_{\beta}e_{\beta}}{\kappa}
{\rm e}^{-\kappa x} . 
\end{equation}
Thus, (\ref{4.18}) becomes
\begin{eqnarray}
n_{\rm qu}^{(\alpha)}(x)& = &\left(1-{\rm e}^{-2x^2/\lambda_{\alpha}^2}\right)
\left[n_{\alpha}(x)+\frac{\sqrt{2}\pi^{3/2}\beta n_{\alpha}e_{\alpha}}{\kappa}
\left(\sum_{\gamma}\lambda_{\gamma}n_{\gamma}e_{\gamma}\right)
{\rm e}^{-\kappa x}\right] \nonumber \\ & &
+ x{\rm e}^{-2x^2/\lambda_{\alpha}^2}n'_{\alpha}(x)+O(\lambda^2). \label{5.6}
\end{eqnarray}
The first line of (\ref{5.6}) is identical to the result (1.14) in Aqua and
Cornu \cite{Aqua04}. 
However, they do not have the second line, since this second line is of 
higher order in $\kappa\beta e^2$ and $\kappa\lambda$ than in the regime 
they have considered.
More generally, terms of higher order in $\lambda$ are also
of higher order in the classical dimensionless coupling parameter
$\kappa\beta e^2$ and,
{\em in this weak coupling regime considered by Aqua and Cornu},
there are no terms $o(\lambda)$ in the density profiles. 

\subsection{Exactly solvable two-dimensional one-component plasma}
The classical two-dimensional one component plasma is 
an exactly solvable model when $\beta e^2=2$. 
In particular, for the semi-infinite geometry, the density profile and 
the two-body densities are exactly known \cite{Jancovici82a}. 
Now, the density profile
\begin{equation} \label{5.7}
n(x)=n\frac{2}{\sqrt{\pi}}\int_0^{\infty}\frac{\exp[-(t-x\sqrt{2})^2]}
{1+\Phi(t)}{\rm d}t,  
\end{equation}
where $\Phi$ is the error function, can be expanded in integer powers of $x$:
all the derivatives of $n(x)$ at $x=0$ are finite (this might be a property of
even integer values of $\beta e^2=2$). 
In particular,
\begin{eqnarray} 
n(0) & = & n\ln 2, \label{5.8} \\
n'(0) & = & -\sqrt{\pi}(2n)^{3/2}\int_0^{\infty}\ln\left(\frac{1+\Phi(t)}{2}
\right){\rm d}t , \label{5.9} 
\end{eqnarray}
where the integral has the numerical value -0.3377. 
The integral which appears in (\ref{4.12}) can be computed: 
\begin{equation} \label{5.10}
\int{\rm d}{\bf r}^{\perp}n^{(2){\rm T}}(0,0;|{\bf r}^{\perp}|)=
-4\sqrt{\frac{2}{\pi}}n^{3/2}\int_0^{\infty}\frac{{\rm e}^{-2t^2}}
{[1+\Phi(t)]^2}{\rm d}t , 
\end{equation}
where the integral on $t$ is related to the previous one: by integrations
per partes, one finds
\begin{equation} \label{5.11}
\int_0^{\infty}\frac{{\rm e}^{-2t^2}}{[1+\Phi(t)]^2}{\rm d}t=
\frac{\sqrt{\pi}}{2}+\frac{\pi}{2}\int_0^{\infty}\ln\left(\frac{1+\Phi(t)}{2}
\right){\rm d}t=0.3558 . 
\end{equation}
The classical surface tension is also known \cite{Russier}: 
\begin{equation} \label{5.12}
\beta\gamma=-\sqrt{\frac{n}{2\pi}}\int_0^{\infty}\ln\left(\frac{1+\Phi(t)}{2}
\right){\rm d}t .   
\end{equation}
Using (\ref{5.12}), (\ref{5.8}), (\ref{5.9}), and (\ref{5.10}) in (\ref{4.12})
gives the surface tension.

\section{Conclusion}
For nearly classical quantum fluids (nearly classical means that the quantum
effects are weak), the Wigner-Kirkwood formalism can be generalized to
semi-infinite fluids, i.e. to fluids confined to a half-space $x\geq 0$ by a
hard plane wall at $x=0$: the boundary condition for the wave functions is that
they have to vanish at $x=0$. 
Under the condition that the classical density profile has all 
its derivatives with respect to $x$ finite at $x=0$, the
surface tension can be expanded in powers, including odd powers, of the
thermal de Broglie wavelength $\lambda$; this is to be compared with the case
of infinite homogeneous fluids with sufficiently smooth interactions, which
have a free energy the expansion of which contains only even powers of
$\lambda$. 
The one-body and many-body densities are more complicated. Their
expansions contain not only powers of $\lambda$, but also non-analytic
position-dependent terms, localized near the wall, of the types
$\exp(-2x^2/\lambda^2)$, $x\exp(-2x^2/\lambda^2)$, etc.

The above assumption of the classical density profile having all its
derivatives with respect to $x$ finite at $x=0$ is not always fulfilled; 
in particular, it fails for Coulomb fluids in the high-temperature 
Debye-H\"uckel regime. 
For such cases, the surface tension can be expanded in integer powers of 
$\lambda$ only up to certain order, then singular terms appear, 
for instance of the type $\lambda^n\ln\lambda$.

As a task for the future, one should consider the exchange effects for
the half-space system geometry. 
We anticipate that, for the considered one-component plasma, 
the exchange effects are exponentially small and therefore negligible 
in the nearly classical regime, like in the bulk case \cite{Jancovici78}.

The extension of the present formalism to more general boundary conditions 
is needed.
For example, in the case of the Neumann boundary condition at $x=0$ 
one has to write down the Boltzmann density in the basis of waves 
$\sqrt{2}\cos(p^x x/\hbar) \exp({\rm i}{\bf p}^{\perp}\cdot 
{\bf r}^{\perp}/\hbar)$ which derivation with respect to $x$
vanishes at $x=0$.
Other geometries of the confining domain (e.g., a strip of finite width) 
are also of interest.

As concerns the model of the one-component plasma, we plan to extend 
the study to the hard walls made of conducting and dielectric materials; 
such walls imply additional image forces acting on charged particles.
In connection with Coulomb fluids, one should also check explicitly 
all available quantum sum rules for the half-space geometry.
 
\ack
We gratefully acknowledge the support received from the European Science 
Foundation (ESF ``Methods of Integrable Systems, Geometry, 
Applied Mathematics''). Partial supports from the CNRS-SAS agreement 
N$^{\rm o}$ 18194 and from a VEGA grant of the Slovak Grant Agency are also
acknowledged.

\end{document}